\documentstyle[12pt,epsfig,amsmath,amssymb,pstricks]{article}
\begin{document}

\title{\bf Quantum mechanics from five physical assumptions}

\author{{Chris Fields}\\ \\
{\it 21 Rue des Lavandi\`eres}\\
{\it Caunes Minervois 11160 France}\\ \\
{fieldsres@gmail.com}}
\maketitle

\begin{abstract}
Five physical assumptions are proposed that together entail the general qualitative results, including the Born rule, of non-relativistic quantum mechanics by physical and information-theoretic reasoning alone.  Two of these assumptions concern fundamental symmetries of physical interactions.  The third concerns the Hilbert-space dimensions and the fourth and fifth the self-interaction Hamiltonians of the systems that function as ``observers'' within the theory.  These assumptions are shown to provide a sufficient motivation for the usual Hilbert-space formalism, and to obviate the observation-related axioms and most interpretative assumptions with which minimal quantum mechanics is typically supplemented.
\end{abstract}

\textbf{Keywords:}  Observation, Quantum foundations, Measurement problem, Born rule, Decoherence, Automata

\section{Introduction}

Quantum mechanics is traditionally presented as an abstract mathematical theory, the correctness of which is evident from over a century of experimentation, but not obvious from its fundamental assumptions.  One of the primary reasons for this non-obviousness is  that it is not clear from the fundamental assumptions what the theory is about; in particular, it is not clear to what the fundamental concept of the state vector refers, and it is not clear how classical information is extracted from quantum systems by measurement.  A wide variety of interpretative additions to the theory have been proposed to address these questions (for reviews see \cite{styer02, schloss04, schloss07}), but none have proven sufficiently satisfactory to gain wide acceptance.  While the dominant response of both the research community and pedagogy to questions of interpretation has been to ``shut up and calculate'', an alternative foundation for quantum mechanics involving clear physical principles that motivate the choice of the Hilbert-space formalism in the way that classical mechanics motivates Lagrangians and Hamiltonians has remained a goal \cite{fuchs02}.

The present paper offers five physical assumptions from which the most general consequences of non-relativistic quantum mechanics, including information loss due to decoherence, non-commuting observables represented as POVMs, and the Born rule can be derived by physical and information-theoretic reasoning with a minimaum of formalism.  Like the mathematical assumptions of traditional, formal presentations of quantum mechanics (henceforth ``TFQM''), these assumptions supplement a pre-theoretic assumption that the universe consists of degrees of freedom that interact in space and time.  The five assumptions are:

\begin{description}
\item[1. Time symmetry:]Physical interactions do not depend on the direction of time.
\item[2. Decompositional equivalence:]Physical interactions do not depend on the specification of boundaries that decompose the universe into systems.
\item[3. Small observers:]The number of degrees of freedom contained within any observer is much smaller than the number of degrees of freedom of that observer's environment.
\item[4. Reliable memory:]The memories of observers are reliable enough to be considered classical.
\item[5. Shared observables:]Observers share observables and hence can confirm or disconfirm each others' observations.
\end{description}

The first two of these assumptions concern fundamental symmetries of physical interactions.  The last three concern the information-coding capacities (corresponding to Hilbert-space dimensions) and functional architectures (corresponding to self-interaction Hamiltonians) of finite physical systems that function as ``observers'' within the theory.  This emphasis on the observer may seem surprising, as with the notable exception of the ``many minds'' interpretation \cite{zeh2000} TFQM generally attributes no particular internal structure to observers, and in some formulations (e.g. \cite{goldstein98}) does without them altogether.  As will be shown below, however, these assumptions have significant and precise physical consequences, and obviate many of the interpretative assumptions with which TFQM is commonly supplemented, including those associated with ``collapse'', ``branching'', and einselection.

In what follows, these assumptions are first presented and motivated, and their relationship to common assumptions or conclusions of TFQM examined.  It is shown that they allow an unambiguous definition of ``observation'' as a classical task, and that this definition together with classical automata theory allows an unambiguous definition of a minimal observer.  A natural definition of quantum-mechanical observables as well-defined POVMs follows, as do standard results concerning commutativity.  It is shown that observations employing these POVMs yield results in accordance with the Born rule.  The given assumptions provide a physical motivation, therefore, for the usual Hilbert-space formalism of TFQM, with interactions described as Hamiltonians.  It is shown that the given assumptions largely obviate the additional axioms and interpretational constructs with which TFQM has traditionally been supplemented.

\section{Five physical assumptions}

\subsection{Time symmetry}

It is assumed that all physical interactions are time-symmetric.  Hence any configuration of degrees of freedom that can be created by physical dynamics can be destroyed by physical dynamics, and vice versa.  In a universe in which configurations of degrees of freedom change, therefore, no configuration can be assumed to be permanent.  Time symmetry is enforced in TFQM by the assumption of a unitary propagator; its assumption is, therefore, fully consistent with traditional presentations.  

\subsection{Decompositional equivalence}

It is assumed that physical interactions do not depend on the specification of boundaries that divide collections of degrees of freedom into ``systems''.  All divisions of the universe into systems are, therefore, entirely arbitrary.  

While the \textit{de facto} practice of physics has always assumed the freedom to select as the ``system of interest'' any collection of degrees of freedom whatever, the assumption of decompositional equivalence presents a \textit{prima facie} conflict with TFQM.  The dependence of TFQM on a non-arbitrary division of the universe into systems has been emphasized by Zurek, who states ``a compelling explanation of what the systems are - how to define them given, say, the overall Hamiltonian in some suitably large Hilbert space - would undoubtedly be most useful'' (\cite{zurek98rough}, p. 1818) and in the absence of such a definition adopts as ``axiom(o)'' of quantum mechanics that ``[T]he Universe consists of systems,'' pointing out that without this additional axiom, ``questions about outcomes cannot be even posed'' (\cite{zurek07grand}, p. 3; see also \cite{schloss04, zurek03rev}).  It is shown below that this conflict between practice and theory results from implicit assumptions of most interpretations of TFQM regarding observers, and that with a consistent treatment of observers ``axiom(o)'' is unnecessary.

\subsection{Small observers}

It is assumed that the configurations of degrees of freedom that function as ``observers'' within the theory are small with respect to their environments.  In particular, it is assumed that the classical information coding capacity of any observer is much smaller than the classical coding capacity of that observer's environment.

The assumption of decompositional equivalence prevents the specification of observer - environment boundaries and hence interactions from first principles; observers can, therefore, only be specified \textit{ad hoc}.  For the purposes of the theory, an ``observer'' is defined as any system capable of carrying out the \textit{task} of observation as characterized in detail below.  As will be shown, typical laboratory data collection systems running on typical laboratory computers satisfy this definition of an observer.

While they make no specific assumptions regarding the internal structures of observers other than that they can ``readily consult the content of their memory'' (\cite{zurek03rev}, p. 759), the ``environment as witness'' \cite{zurek04, zurek05} and quantum Darwinism \cite{zurek07grand, zurek06, zurek09rev} programs assume that observers are much smaller than their environments.  As will be seen below, the qualitative ``picture'' of observation derived from the present assumptions is very similar to that of quantum Darwinism, although the specific assumptions of quantum Darwinism, including einselection, are not required.

\subsection{Reliable memories}

It is assumed that observers have memories that can be considered classical.  As will be shown below, this assumption is required to define the task of observation.

The assumption of reliable memory introduces a second \textit{prima facie} conflict with TFQM.  Interpretations of TFQM in the von Neumann - Everett tradition characterize the observer as a quantum system, the final state of which is determined by entanglement with the system observed as represented by the von Neumann chain:

\begin{equation}
(\sum_{k}\lambda_{k}|s_{k}\rangle)|\mathcal{A}^{\mathit{i}}\rangle|\mathcal{O}^{\mathit{r}}\rangle\,\rightarrow \mathit{\sum_{k}\lambda_{k}|s_{k}\rangle|a_{k}\rangle|o_{k}\rangle\,\rightarrow |s_{f}\rangle|a_{f}\rangle|o_{f}\rangle}
\end{equation}

where $|\mathcal{A}^{\mathit{i}}\rangle$ and $|\mathcal{O}^{\mathit{r}}\rangle$ refer to the unmeasured initial state of an apparatus $\mathcal{A}$ and ``ready'' state of the observer $\mathcal{O}$, the microscopic quantum system being observed is described by basis vectors $|s_{k}\rangle$, and $|s_{f}\rangle|a_{f}\rangle|o_{f}\rangle$ is the joint final state.  Quantum systems cannot be cloned \cite{wooters82}, so the final observer state $|o_{f}\rangle$ cannot serve as a recallable memory state.  This issue is often finessed (e.g. Wigner's friend does not question Wigner's memory), but it can only be eliminated by either eliminating observers altogether (e.g. \cite{goldstein98}) or invoking decoherence either at the observer-environment boundary (e.g. \cite{zurek03rev}) or internal to the observer (e.g. \cite{tegmark00}).  

\subsection{Shared observables}

It is assumed that observers share observables, i.e. that they can observe the ``same things.''  Such an assumption is required for observations to be confirmable or disconfirmable, and hence for science to be possible.  

Observables are defined independently of observers in most interpretations of TFQM, so the question of whether they are shared does not arise.  However, if quantum states are interpreted as purely subjective, i.e. as descriptions of observers' beliefs or knowledge (e.g. \cite{fuchs02}), the question of whether observations are confirmable is typically unresolved.

\section{Minimal observers}

This section first characterizes observation as a \textit{task} in classical terms.  It then employs classical automata theory to define the minimal classical observer capable of carrying out this task.  The classical assumptions are then relaxed, and it is shown that the functional architecture of the minimal observer is alone sufficient to enforce compliance with LOCC.  The section closes by deriving an observer-relative analog of the no-cloning theorem of TFQM.

Consider a laboratory environment containing $N$ discrete, macroscopic items of apparatus $\mathcal{A}_{\mathit{i}}$, each of which has a macroscopic ``read-out'' component $\mathcal{R}_{\mathit{i}}$ comprising, for example, a finite collection gauges or displays.  Any read-out $\mathcal{R}_{\mathit{i}}$ is capable of indicating any one of $k$ blocks of classical information that can be regarded, without loss of generality, as encoded by $k$ finite-precision real values $r_{ik}$.  An \textit{observer} is any finite physical system capable of 1) reliably re-identifying each of the $\mathcal{A}_{\mathit{i}}$ across an extended period of time during which each of the $\mathcal{A}_{\mathit{i}}$ undergo multiple state transitions; 2) reliably re-identifying each of the $\mathcal{R}_{\mathit{i}}$ across multiple state transitions of the relevant $\mathcal{A}_{\mathit{i}}$ and reliably recording its value $r_{ik}$ on each observation; 3) reliably reporting the accumulated results as a sequence of values $r_{ik}$, each tagged by its source, at the end of the observing session.  Assuming that an observer $\mathcal{O}$ begins each observation in a ``ready'' state $|\mathcal{O}^{\mathit{r}}\rangle$, the task of observation can be summarized by the functional decomposition shown in Fig. 1.  

\psset{xunit=1cm,yunit=1cm}
\begin{pspicture}(0,0)(16,16)
\put(14.5,15.1){\vector(-1,0){10.2}}
\put(3.2,15){$|\mathcal{O}^{\mathit{r}}\rangle$}
\put(3.5,14.7){\vector(0,-1){0.7}}
\pspolygon(2,13)(3.5,14)(5,13)(3.5,12)
\put(2.5,12.9){observing?}
\put(5,13){\vector(1,0){4.5}}
\put(5.5,13.2){No}
\put(3.5,12){\vector(0,-1){1}}
\pspolygon(2,10)(3.5,11)(5,10)(3.5,9)
\put(2.5,9.9){$\mathcal{A}_{\mathit{1}}$ signal?}
\put(5,10){\line(1,0){1.5}}
\put(5.5,10.2){No}
\put(6.5,10){\line(0,-1){.3}}
\psdot(6.5,9.3)
\psdot(6.5,9)
\psdot(6.5,8.7)
\put(6.5,8.3){\vector(0,-1){.3}}
\pspolygon(5,7)(6.5,8)(8,7)(6.5,6)
\put(5.5,6.9){$\mathcal{A}_{\mathit{N}}$ signal?}
\psdot(14.5,7)
\put(8,7){\vector(1,0){6.5}}
\put(8.5,7.2){No}
\pspolygon(9.5,12.5)(9.5,13.5)(12.5,13.5)(12.5,12.5)
\put(9.7,12.9){Report records}
\put(11,12.5){\vector(0,-1){1.5}}
\pspolygon(9.5,10)(9.5,11)(12.5,11)(12.5,10)
\put(9.8,10.4){Flush records}
\put(11,10){\line(0,-1){.5}}
\psdot(14.5,9.5)
\put(11,9.5){\vector(1,0){3.5}}
\put(3.5,9){\vector(0,-1){0.5}}
\pspolygon(2,7.5)(2,8.5)(5,8.5)(5,7.5)
\put(2.5,7.9){Extract $\mathcal{R}_{\mathit{1}}$}
\put(3.5,7.5){\vector(0,-1){5}}
\pspolygon(2,1.5)(2,2.5)(5,2.5)(5,1.5)
\put(2.5,1.9){Record $\mathcal{R}_{\mathit{1}}$}
\put(3.5,1.5){\line(0,-1){0.2}}
\put(3.5,1.3){\line(1,0){11}}
\put(14.5,1.3){\line(0,1){13.8}}
\psdot(9.5,2.2)
\psdot(9.5,1.9)
\psdot(9.5,1.6)
\put(6.5,6){\vector(0,-1){.5}}
\pspolygon(5,4.5)(5,5.5)(8,5.5)(8,4.5)
\put(5.5,4.9){Extract $\mathcal{R}_{\mathit{N}}$}
\put(6.5,4.5){\vector(0,-1){0.5}}
\pspolygon(5,3)(5,4)(8,4)(8,3)
\put(5.5,3.4){Record $\mathcal{R}_{\mathit{N}}$}
\put(6.5,3){\line(0,-1){0.5}}
\psdot(14.5,2.5)
\put(6.5,2.5){\vector(1,0){8}}

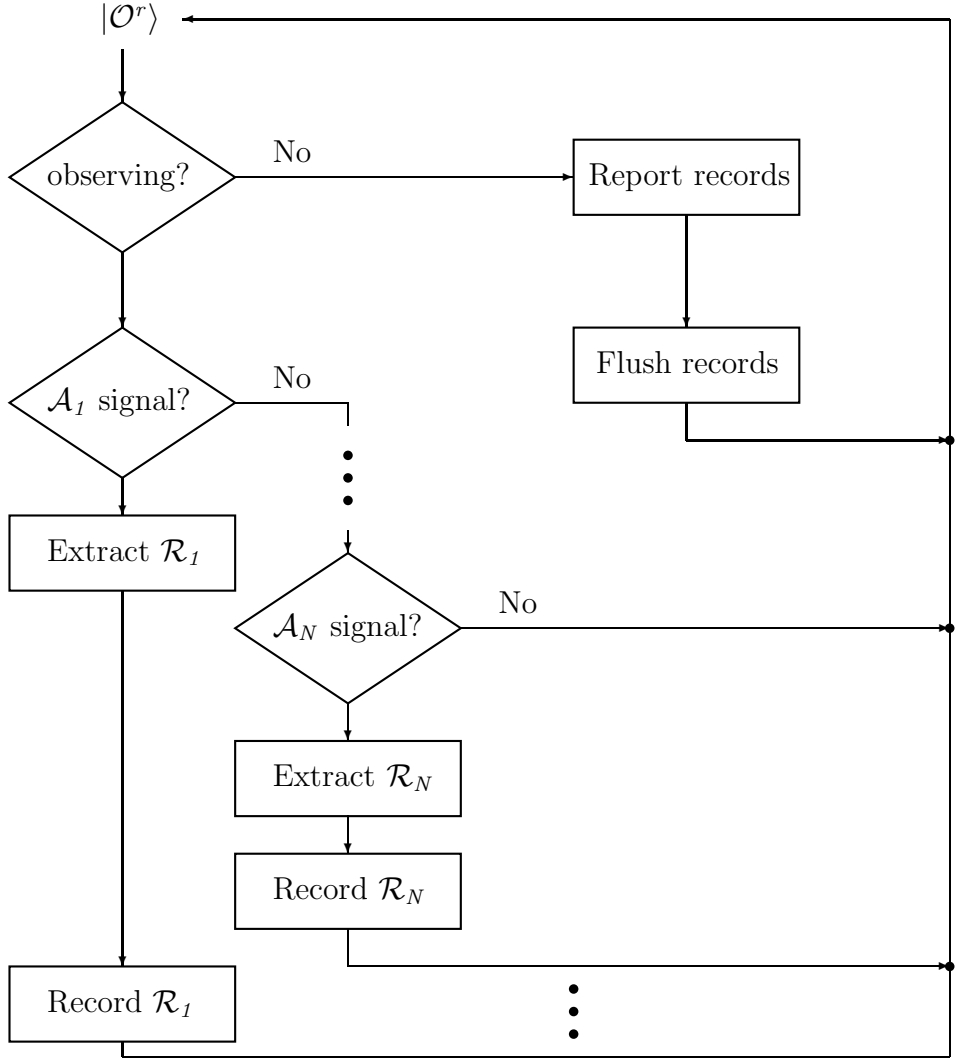
\put(2.5,0.5){\textit{Fig. 1: Functional decomposition of multiple-observation task}}
\end{pspicture}

Intuitively, any observer capable of carrying out the task shown in Fig. 1 must \textit{know how} to identify the $\mathcal{A}_{\mathit{i}}$ and \textit{know how} to extract the values $r_{ik}$ from the $\mathcal{R}_{\mathit{i}}$.  These intuitions about the prior knowledge of observers are formalized by classical automata theory.  Suppose $\mathcal{O}$ interacts with the $\mathcal{A}_{\mathit{i}}$ by looking at them, i.e. by means of scattered ambient photons.  The state of the environment $\mathcal{E}$ shared by $\mathcal{O}$ and the $\mathcal{A}_{\mathit{i}}$ then serves as a classical communication channel.  Represent the signals transmitted along this channel as a sequence of classical bits, and assume that $\mathcal{O}$ receives these bits without noise or overlap.  The observation task then consists of 1) identifying the bit strings that encode well-formed messages from each source $\mathcal{A}_{\mathit{i}}$; 2) extracting the component of each of these well-formed bit strings that encodes the read-out value $r_{ik}$; 3) writing these read-out values to memory, tagged by source.  A well-formed message clearly cannot consist \textit{only} of a read-out value (e.g. ``2''), as it must also include bits that identify the source.  Classical automata theory demonstrates that $\mathcal{O}$ cannot define, from any finite sample of input from $\mathcal{E}$, either the bit patterns that distinguish messages by source or which bits within each message encode the read-out values $r_{ik}$ to be recorded, even if $\mathcal{O}$ is permitted finite diagnostic inputs to the $\mathcal{A}_{\mathit{i}}$ \cite{moore56}.  If $\mathcal{O}$ cannot infer the rules necessary to recognize the $\mathcal{A}_{\mathit{i}}$ or extract the $r_{ik}$ from a finite sample of the input stream, $\mathcal{O}$ must encode them in advance.  As these signal-source recognition and read-out value extraction rules are needed on every cycle of observation, they must be encoded in $\mathcal{O}$'s ready state $|\mathcal{O}^{\mathit{r}}\rangle$.  This encoding must, moreover, be reliable; hence the assumption that observers have reliable memories is required to define the observation task.

Successful completion of the observation task requires not just identification of the $\mathcal{A}_{\mathit{i}}$ and extraction of the $r_{ik}$, it requires doing so on multiple cycles, and then reporting the accumulated results.  For a fixed, finite set of $\mathcal{A}_{\mathit{i}}$ and a fixed number of observation cyles, a classical finite-state machine with a fixed memory will meet these requirements.  However, a general observer capable of carrying out the task of Fig. 1 for any observation period and any possible finite set of $\mathcal{A}_{\mathit{i}}$ cannot be assumed to pre-allocate memory for all possible observation periods or to encode recognizers for all possible $\mathcal{A}_{\mathit{i}}$; such an observer must, therefore, incorporate a dynamically-allocatable memory and a capability to acquire and store specifications for recognizers of new $\mathcal{A}_{\mathit{i}}$ not previously encountered.  These capabilities require an architecture functionally equivalent to a classical Turing Machine. 

A \textit{classical minimal observer} is a finite physical system capable of performing the task summarized in Fig. 1 for arbitrary finite observation times and any specified finite set of $\mathcal{A}_{\mathit{i}}$.  The above considerations show that a classical minimal observer must implement a Turing-equivalent functional architecture and must encode signal-source recognizers for the $\mathcal{A}_{\mathit{i}}$ and read-out value extractors for the $r_{ik}$ in its ready state $|\mathcal{O}^{\mathit{r}}\rangle$.  Classical minimal observers are commonplace; any laboratory data-collection system running on a general-purpose computer is a classical minimal observer.

Having defined a classical minimal observer, the assumptions employed to describe the observation task in classical terms can be relaxed.  Three assumptions are involved.  First, the signals transmitted through $\mathcal{E}$ were assumed to be strings of discrete bits, received in the absence of noise and without overlap.  This assumption conflicts with the intuitions that physical dynamics are continuous and environments are noisy; hence it is relaxed to an assumption of arbitrary continuous input to $\mathcal{O}$ from $\mathcal{E}$.  Second, the apparatus $\mathcal{A}_{\mathit{i}}$ were assumed to be discrete entities, and it was implicitly assumed that only signals from the $\mathcal{A}_{\mathit{i}}$ were transmitted through $\mathcal{E}$.  By treating the boundaries of the $\mathcal{A}_{\mathit{i}}$ as ``real'', this assumption straightforwardly conflicts with decompositional equivalence.  It is therefore relaxed to an assumption consistent with decompositional equivalence, namely that signals from all possible collections of degrees of freedom, including all single degrees of freedom, are simultaneously propagated through $\mathcal{E}$.  Third, it was implicitly assumed that the $\mathcal{O-E}$ boundary and hence the $\mathcal{O-E}$ interaction were well-defined.  This assumption also conflicts with decompositional equivalence.  To maintain consistency, an \textit{ad hoc} $\mathcal{O-E}$ boundary that is small enough to comply with the assumption of small observers but large enough to contain the degrees of freedom that implement $\mathcal{O}$'s functional architecture is assumed; such a boundary can be drawn, for example, at the visible surface of $\mathcal{O}$.  This \textit{ad hoc} boundary allows the definition of an \textit{ad hoc} $\mathcal{O-E}$ interaction, which will be denoted $H_{\mathcal{O-E}}$ in anticipation of the adoption of a state representation in which interactions can be represented by Hamiltonians.  It also allows the definition of a self interaction $H_{\mathcal{O-O}}$.  As the definition of observation involves the assumption that $\mathcal{O}$ continues to exist as an identifiable entity, the \textit{ad hoc} interactions $H_{\mathcal{O-E}}$ and $H_{\mathcal{O-O}}$ must at least approximately commute.

The task of observation can now be described using these relaxed assumptions.  The \textit{ad hoc} observer $\mathcal{O}$ interacts via $H_{\mathcal{O-E}}$ with $\mathcal{E}$.  The environment $\mathcal{E}$ encodes in its continuous dynamics signals from all possible collections of degrees of freedom, including those previously but no longer singled out and labeled as the $\mathcal{A}_{\mathit{i}}$.  The task of $\mathcal{O}$ is to interact with $\mathcal{E}$ over an extended period of time, and at the end of that period to report a set of source-tagged real values $r_{ik}$ indicating the states of the no longer specified read-outs $\mathcal{R}_{\mathit{i}}$.  As before, $\mathcal{O}$ is capable of completing this task if, but only if, $\mathcal{O}$ implements a Turing-equivalent architecture with reliable memory, and encodes in $|\mathcal{O}^{\mathit{r}}\rangle$ executable specifications of the signal-source recognizers required to identify signals from the $\mathcal{A}_{\mathit{i}}$ and the read-out extractors required to extract the $r_{ik}$.  None of the functional requirements on $\mathcal{O}$ change as a result of the re-description of observation with these relaxed assumptions, because nothing about the \textit{physical situation} with which $\mathcal{O}$ is faced changes as a result of the re-description.  Only the information included in the theory changes; the information and architecture that $\mathcal{O}$ must possess to perform the task of observation while immersed in $\mathcal{E}$ do not.  A classical minimal observer is, therefore, a \textit{minimal observer} under the relaxed conditions.  This result can be formalized as a theorem:

\textit{Theorem} (\textbf{Enforced LOCC}):  \textit{Ad hoc} specification of a minimal observer by itself enforces LOCC.

\textit{Proof:}  A minimal observer $\mathcal{O}$ interacts directly only with the environment $\mathcal{E}$, and only at the specified \textit{ad hoc} $\mathcal{O-E}$ boundary.  Hence all observations are local.  A minimal observer requires a memory reliable enough to be considered classical.  Minimal observers are finite physical systems, so this classical memory can encode at most a finite number $i$ of executable specifications of signal-source recognizers and a finite number $k$ of executable specifications of extractors of finite-precision values $r_{ik}$ from each signal source.  Communication from any signal source to $\mathcal{O}$ is therefore effectively classical.  $\square$

The assumption of small observers entails a further result that, as will be shown below, renders observations by minimal observers consistent with the predictions of TFQM.

\textit{Theorem} (\textbf{No Replication}):  A minimal observer cannot replicate the initial conditions of $\mathcal{E}$.

\textit{Proof:}  Minimal observers have fewer degrees of freedom than their environments, so the coding capacities of their memories are smaller than the coding capacities of their environments.  Therefore, a minimal observer $\mathcal{O}$ cannot fully represent any state of $\mathcal{E}$ in memory, and in particular cannot fully represent the initial state $|\mathcal{E}^{\mathit{i}}\rangle$ that begins any cycle of observation.  Therefore $\mathcal{O}$ cannot recognize any initial state $|\mathcal{E}^{\mathit{i}}\rangle$.  $\square$

No Replication is an observer-relative analog of the no-cloning theorem of TFQM; it implies that even if an environmental state could be cloned, no observer could recognize it as cloned.

\section{Observables and the Born rule for minimal observers}

Let $\lbrace P_{i}\rbrace$ denote the set of signal-source recognizers and $\lbrace R_{ik}\rbrace$ denote the set of read-out value extractors implemented by a minimal observer $\mathcal{O}$.  These functions are implemented by the self-interaction $H_{\mathcal{O-O}}$ in the sense defined by classical computer science: from given initial states $|\mathcal{O}^{\mathit{r}}\rangle$ and $|\mathcal{E}^{\mathit{i}}\rangle$, particular, sequential changes in the configurations of degrees of freedom of $\mathcal{O}$ can be mapped to sequential abstract steps in the execution of an algorithm specifying one of the $P_{i}$, and further particular, sequential changes in the configurations of degrees of freedom of $\mathcal{O}$ can be mapped to sequential abstract steps in the execution of an algorithm specifying one of the $R_{ik}$ (e.g. \cite{tan76}).  The notations ``$P_{i}$'' and ``$R_{ik}$'' will be employed below to refer either to the abstract recognizers or extractors or to their physical implementations by $H_{\mathcal{O-O}}$ as determined by context.

As shown in Fig. 1, the $R_{ik}$ are effectively orthogonal; if one of the $R_{ik}$ ``fires'' following an input, $\mathcal{O}$ returns to $|\mathcal{O}^{\mathit{r}}\rangle$ and a new cycle of observation is initiated.  Moreover, $\mathcal{O}$'s architecure guarantees both that some $R_{ik}$ will return a real value $r_{ik}$ whenever a signal source is recognized by some $P_{i}$, and that no $R_{ik}$ will return a value in the case of non-recognition (i.e., $\mathcal{O}$ returns to $|\mathcal{O}^{\mathit{r}}\rangle$ with no memory-write operation).  Under these conditions, for any stipulated $H_{\mathcal{O-E}}$ the set $\lbrace O_{ik}\rbrace$, where $O_{ik} = H_{\mathcal{O-E}} \circ P_{i} \circ R_{ik}$, forms an $\mathcal{A}_{\mathit{i}}$-specific POVM over states of $\mathcal{E}$.  The POVM $\lbrace O_{ik}\rbrace$ produces read-out values $r_{ik}$ whenever $\mathcal{A}_{\mathit{i}}$ is recognized by $\mathcal{O}$, and is therefore an $\mathcal{A}_{\mathit{i}}$-specific \textit{observable} for $\mathcal{O}$.

Suppose observers $\mathcal{O}^{\mathit{(1)}}$ and $\mathcal{O}^{\mathit{(2)}}$ implement POVMs $\lbrace O^{(1)}_{ik}\rbrace$ and $\lbrace O^{(2)}_{ik}\rbrace$ respectively, and that:

\begin{equation}
\forall i,k \, \mathrm{and} \, \forall |\mathcal{E}^{\mathit{i}}\rangle, \mathit{O^{(1)}_{ik}}|\mathcal{E}^{\mathit{i}}\rangle = \mathit{O^{(2)}_{ik}}|\mathcal{E}^{\mathit{i}}\rangle = \mathit{r_{ik}}.
\end{equation}

In this case $\lbrace O^{(1)}_{ik}\rbrace$ and $\lbrace O^{(2)}_{ik}\rbrace$ are output-equivalent and $\mathcal{O}^{\mathit{(1)}}$ and $\mathcal{O}^{\mathit{(2)}}$ can be said to share a POVM $\lbrace O_{ik}\rbrace$.  Multiple minimal observers will agree that a collection of degrees of freedom $\mathcal{A}_{\mathit{i}}$ is present in their shared environment if and only if they share a POVM $\lbrace O_{ik}\rbrace$ that outputs a single set $\lbrace r_{ik}\rbrace$ of values for the read-out states of $\mathcal{A}_{\mathit{i}}$.  The notion of ``objectivity'' has been defined within quantum Darwinism as:

\begin{quotation}
``A property of a physical system is \textit{objective} when it is:
\begin{list}{\leftmargin=2em}
\item
1. simultaneously accessible to many observers,
\item
2. who are able to find out what it is without prior knowledge about the system of interest, and 
\item
3. who can arrive at a consensus about it without prior agreement."
\end{list}
\begin{flushright}
(p. 1 of \cite{zurek04}; p. 3 of \cite{zurek05})
\end{flushright}
\end{quotation}

A set of minimal observers that share POVMs will agree, using this definition, that the $\mathcal{A}_{\mathit{i}}$ they jointly observe are ``objective'' and hence effectively classical.

Suppose now that $\mathcal{O}^{\mathit{(1)}}$ and $\mathcal{O}^{\mathit{(2)}}$ share $\mathcal{E}$ and share POVMs $\lbrace O_{ik}\rbrace$ and $\lbrace Q_{jl}\rbrace$.  Suppose $\mathcal{O}^{\mathit{(1)}}$ performs observations with $\lbrace O_{ik}\rbrace$ and then with $\lbrace Q_{jl}\rbrace$, and obtains results $r_{ik}$ and $s_{jl}$ respectively; simultaneously $\mathcal{O}^{\mathit{(2)}}$ performs observations with $\lbrace Q_{jl}\rbrace$ and then with $\lbrace O_{ik}\rbrace$, and obtains results $s^{\prime}_{jl}$ and $r^{\prime}_{ik}$ respectively.  Clearly $\mathcal{O}^{\mathit{(1)}}$ and $\mathcal{O}^{\mathit{(2)}}$ can obtain the same results (i.e. $r_{ik} = r^{\prime}_{ik}$ and $s_{jl} = s^{\prime}_{jl}$) only if the systems $\mathcal{A}_{\mathit{i}}$ and $\mathcal{A}_{\mathit{j}}$ recognized by the shared $P_{i}$ and $P_{j}$ encode signals in non-interacting and hence orthogonal components of $\mathcal{E}$.  Hence:

\textit{Theorem} (\textbf{Commutativity}): Observables $\lbrace O_{ik}\rbrace$ and $\lbrace Q_{jl}\rbrace$ commute only if the identified systems $\mathcal{A}_{\mathit{i}}$ and $\mathcal{A}_{\mathit{j}}$ encode signals in orthogonal components of $\mathcal{E}$.

\textit{Proof:} Suppose otherwise, and let $X$ be the component of $\mathcal{E}$ that encodes signals from both $\mathcal{A}_{\mathit{i}}$ and $\mathcal{A}_{\mathit{j}}$.  The POVM $\lbrace O_{ik}\rbrace$ is a resolution of the identity for $\mathcal{E}$ and is therefore well-defined over $X$ in the absence of $\mathcal{A}_{\mathit{j}}$.  The addition of $\mathcal{A}_{\mathit{j}}$ to $\mathcal{E}$ cannot, therefore, affect the values returned by $\lbrace O_{ik}\rbrace$, contradicting the supposition.  $\square$

This commutativity theorem replicates for the system- and observer-dependent observables defined in the current framework the general commutativity result obtained for system- and observer-independent observables in TFQM.  In the current framework, however, observers can only demonstrate commutativity of observables by cooperative experiments such as described above, since they are forbidden by the No Replication theorem from replicating initial states of $\mathcal{E}$.  As the minimum overlap between components of $\mathcal{E}$ encoding signals from two distinct sources is one degree of freedom, the minimum perturbation of experimental results obtained with non-commuting observables due to signal interference is the perturbation due to altering one degree of freedom of $\mathcal{E}$.   

It follows trivially from the Commutativity theorem that no two observables that pick out the same external collection of degrees of freedom $\mathcal{A}_{\mathit{i}}$ commute.  Hence:

\textit{Theorem} (\textbf{Objective ignorance}): Observers cannot determine the configurations of external degrees of freedom that yield, on observation, any given value $r_{ik}$.

\textit{Proof:}  Any such determination would require an observable $\lbrace Q_{il}\rbrace$ that picked out $\mathcal{A}_{\mathit{i}}$ and assigned values $s_{il}$ to each configuration of its degrees of freedom.  By the Commutatitivity theorem, no such observable would commute with $\lbrace O_{ik}\rbrace$, therefore no such $\lbrace Q_{il}\rbrace$ could be used to determine which configurations of $\mathcal{A}_{\mathit{i}}$ yield which $r_{ik}$.  $\square$

The Objective Ignorance theorem provides an analog in the current framework of an earlier result showing that observers cannot demonstrate redundant environmental encoding of pointer states in TFQM under the assumptions of quantum Darwinism \cite{fields10}.

As shown by Zurek \cite{zurek05env}, provable ignorance generates the Born rule.  Stripped of formalized assumptions about how states are to be described, the Born rule states an absence of bias, both on the part of observers and on the part of the observables employed to carry out observations.  It requires that the probability $P_{k}$ with which an observer $\mathcal{O}$ will report the value $r_{ik}$ following application of an observable $\lbrace O_{ik}\rbrace$ depends only on the actual fraction $F(k)$ of configurations of evironmental degrees of freedom that encode $r_{ik}$, i.e.:  

\textit{Theorem} (\textbf{Born Rule}): $P_{k} = F(k)$.

In the current framework, the only source of ``objectivity'' is agreement among observers; hence $F(k)$ can at best be estimated by an objectively ignorant observer employing envariance.  Otherwise the proof follows that given by Zurek \cite{zurek05env}.

\textit{Proof}:
Let $C = \bigcup C_{i}$ be an ancilla into which an observer $\mathcal{O}^{\mathit{(1)}}$ sorts configurations $|e_{j}\rangle$ of environmental degrees of freedom using an observable $\lbrace O_{ik}\rbrace$.  If $\lbrace O_{ik}\rbrace |e_{j}\rangle$ yields a value $r_{ik}$, $|e_{j}\rangle$ is sorted into $C_{k}$; if $\lbrace O_{ik}\rbrace |e_{j}\rangle$ is NULL, $|e_{j}\rangle$ is discarded.  Because $\mathcal{O}^{\mathit{(1)}}$ is objectively ignorant, this sorting process, carried on long enough, will produce a $C$ for which the fractions $F(k)$ are arbitrarily representative of their values in $\mathcal{E}$.  Now suppose a second observer $\mathcal{O}^{\mathit{(2)}}$ that shares $\lbrace O_{ik}\rbrace$ examines the configurations in $C$ sequentially.  Because $\mathcal{O}^{\mathit{(2)}}$ is objectively ignorant, $\mathcal{O}^{\mathit{(2)}}$'s sequential selections of configurations cannot be biased for $k$, so the probability $P_{k}$ that the next configuration $|e_{j}\rangle$ selected will be such that $\lbrace O_{ik}\rbrace |e_{j}\rangle = r_{ik}$ is $F(k)$ as required.  $\square$

\section{Choice of formalism}

The previous section shows that the general qualitative results of TFQM can be derived from the five physical assumptions stated in the Introduction by physical and information-theoretic reasoning alone, without any assumptions beyond the notion of a ``configuration'' and the definition of a POVM regarding the formal descriptions of either states or operators.  What is needed in practice, however, is a formalism that permits calculation.  It is therefore reasonable to ask what mathematical assumptions would yield a formalism adequate to the description of the physical results already obtained.  Such a formalism would ideally enforce, or at least strongly suggest, time symmetry, decompositional equivalence, and objective ignorance of states following operations represented by POVMs.

As is well known, the formalism of \textit{minimal} TFQM, without observation-related axioms or supplementary interpretative assumptions, satisfies these criteria.  Time symmetry is enforced by a unitary propagator; additional assumptions, such as irreversible information ``loss'' due to decoherence \cite{schloss07}, must be imposed to break this time symmetry.  Decompositional equivalence is enforced by linearity; additional assumptions, such as Zurek's ``axiom(o)'' that systems with well-defined boundaries exist \cite{zurek07grand}, must be introduced to break this symmetry.  Objective ignorance on the part of observers is enforced by the concepts of open systems and the universal state vector; additional assumptions, such as the supplementary axioms that isolated systems can somehow be ``prepared'' or that observations can be ``repeated'' are required to circumvent this ignorance.

Interposing a structured minimal observer with encoded information specifying what is to be observed between the state vector defined by minimal TFQM and an observational outcome, as is always the case in practice, obviates the observation-oriented axioms and most of the interpretative assumptions with which minimal TFQM is commonly supplemented.  Minimal observers cannot help but identify particular external systems as sources of signals; neither decoherence nor einselection nor additional axioms are needed to define ``systems.''  Minimal observers cannot help but report discrete real values as observational outcomes; notions of state-vector ``collapse'' or ``branching'' are unnecessary, as are the ``guiding fields'' of Bohmian mechanics.  No observable systems are isolated, initial states cannot be ``prepared'', and nothing assures that repeated observations will yield the same outcome.  Minimal observers agree about ``objectivity'' if and only if they share POVMs; quantum Darwinism is not required to assure that they do.  Minimal observers obey the Born rule in consequence of their functional architectures; it does have to be assumed as an axiom.  Minimal observers have no more psychology than laboratory computers; no assumptions about ``consciousness,'' ``experience'' or ``belief revision'' are required. 

The five physical assumptions presented here, together with a representation of states as vectors in a Hilbert space and interactions as Hamiltonians acting on such states, appear to constitute an adequate, self-consistent, and relatively unproblematic foundation for non-relativistic quantum mechanics.  The physical assumptions are straightforward and pose no particular challenges to intuition; the three assumptions regarding observers, in particular, accord completely with ordinary experience.  Unlike in TFQM, the assumption of the Hilbert space formalism in this conext is well-motivated by simple physical reasoning.  The proposed foundation is not entirely free of philosophical issues; while a realist interpretation of the environment and its state is straightforward, it is difficult to escape an observer-dependent, constructivist view of the ``objects'' of ordinary experience.  This philosophical issue is, however, not unique to quantum mechanics, and seems a small price to pay for freedom from such issues as collapsing wave functions, infinitely-branching trees of possible outcomes, or axiomatic postulates that particular systems exist.

\section{Conclusion}

Since the pioneering work of von Neumann \cite{vonN32}, the observer of a quantum state $|S\rangle$ has been treated as a physical system that becomes entangled with $|S\rangle$.  The fact that observers report definite outcomes of experiments has, therefore, been a mystery.  Explanations of this mystery have supplemented quantum mechanics with a wide variety of additional assumptions, but have not questioned the fundamental premise of system - observer entanglement.  

The foundational assumptions for quantum mechanics proposed here depart from this tradition by treating ``observer'' not as the name of a system, but as a functional requirement.  An ``observer'' in this treatment \textit{must} report a definite outcome; this is a requirement of the \textit{task} of observation.  Treating observation as a functional requirement naturally leads to the concept of a minimal observer, a concept fully formed by classical automata theory over 50 years ago.  A minimal observer functions in a quantum environment exactly as would be expected for a system with finite observational and memory resources: initial states cannot be fully characterized, observations are restricted by commutativity requirements on observables, and results are reported with probabilities given by the Born rule.  Minimal observers behave, in other words, like laboratory computers or graduate students, not like all-knowing oracles capable of perceiving mixtures between dead and living cats.

Zurek \cite{zurek03rev} proposed that the world appears ``quantum'' because ``Hilbert space is big'';  Fuchs \cite{fuchs02} suggests that the Hilbert-space dimension of a physical system determines its ``sensitivity to touch'' and hence the ``zing'' it can be expected to display following an experimental intervention.  The present work shows that quantum mechanics follows from the assumption of a big Hilbert space, without fixed boundaries to separate systems of interest, with which a pragmatic observer with limited resources is nonetheless required to cope.

\end{document}